\documentclass[journal]{IEEEtran}
\usepackage{cite,url,subfigure,epsfig,graphicx}
\usepackage{tipa}
\usepackage{makecell}
\usepackage{multirow}
\usepackage{bbm}
\usepackage{mathrsfs}
\usepackage{array}
\usepackage{amsfonts}
\usepackage{cite,url,subfigure,epsfig,graphicx}
\usepackage{amssymb,amsmath}
\usepackage{indentfirst}
\usepackage{pifont}
\usepackage{algorithmic}
\usepackage{algorithm}
\usepackage{cases}
\usepackage{soul}
\usepackage{booktabs}
\usepackage[table]{xcolor}
\usepackage{graphicx}
\IEEEoverridecommandlockouts
\makeatletter
\usepackage{times,verbatim,amsfonts,amsmath,color}
\begin{document}

\title{Secure UAV Swarms in Low-Altitude Wireless Networks: Challenges and Solutions}

\author{
\IEEEauthorblockN{Yuntao~Wang, Haojia~Yang, Han~Liu, Jianle~Ba, and Zhou~Su\IEEEauthorrefmark{1}}\\
\IEEEauthorblockA{
School of Cyber Science and Engineering, Xi'an Jiaotong University, China\\
\IEEEauthorrefmark{1}Corresponding author: zhousu@ieee.org
}}
\maketitle

\begin{abstract}
Unmanned aerial vehicle (UAV) swarms are increasingly deployed in vast low-altitude applications, owing to their capabilities in distributed sensing, flexible communication, and autonomous coordination. Nevertheless, the open and highly dynamic operating environment of UAV swarms introduces serious security risks, including GPS spoofing, insider threats, and multi-hop intrusion. These threats are aggravated by limited on-board resources, frequently changing network topology, and the presence of intelligent adversaries. 
To tackle these issues, this paper proposes a cloud-edge-end collaborative defense framework for UAV swarms. Based on this framework, three complementary mechanisms are developed. First, a cooperative perception scheme is designed to resist GPS spoofing via interactive attack-defense game modeling. Second, a behavior-driven authentication method with trust evaluation is developed to mitigate insider threats. Third, a multi-agent attack forensics framework is devised to intelligently trace the propagation paths of multi-hop attacks in UAV networks. Experimental results validate the effectiveness of the proposed approaches. Finally, several open research directions are outlined.
\end{abstract}

\IEEEpeerreviewmaketitle

\section{Introduction}
\IEEEPARstart{W}{ith} the deep integration of intelligent unmanned systems and low-altitude wireless networks, unmanned aerial vehicle (UAV) swarms are rapidly evolving from isolated aerial platforms into complex cyber-physical systems that tightly integrate sensing, communication, decision-making, and execution \cite{Jin2025Codesign}. Nowadays, low-altitude airspace is transitioning from sparsely utilized environments to highly dynamic and densely populated networked systems. 
According to a recent report\footnote{https://www.marketreportsworld.com/market-reports/drone-market-14714844}, more than 32.5 million UAVs were operational worldwide as of 2024, while the global UAV market is projected to increase from USD 16.37 billion in 2026 to USD 24.12 billion by 2035. 

Enabled by distributed intelligence and autonomous multi-agent collaboration, UAV swarms have demonstrated transformative potential across a wide range of mission-critical applications \cite{Du2025Survey}, including wide-area reconnaissance, low-altitude logistics, infrastructure inspection, public safety surveillance, and emergency response. 
For instance, UAV swarm operations supported by satellite communication systems such as Starlink have demonstrated unprecedented capabilities in distributed sensing, resilient communication, and coordinated mission execution in complex environments \cite{Li2026Securing}. These developments impose increasingly stringent requirements on secure collaboration, resilient networking, and autonomous defense capabilities in UAV swarms.

However, the distributed, autonomous, and highly dynamic nature of UAV swarms also introduces fundamentally new security challenges, making them vulnerable to both sophisticated external attacks and internal compromises \cite{Li2026Securing}. Specifically, a series of new security challenges may arise in low-altitude wireless networks.

\textit{1) GPS Spoofing:} GPS signals are widely used for UAV navigation and are inherently weak and openly broadcast, making them vulnerable to spoofing attacks \cite{Sathaye2022Experimental}. Once compromised, navigation deviations may propagate through inter-UAV coordination, potentially leading to large-scale swarm mission failure. To mitigate this threat, conventional defense approaches typically rely on multi-source sensor fusion on individual UAVs. However, due to limited sensing redundancy and onboard resources, individual UAV-based defenses remain insufficient, necessitating low-cost swarm-level defense to ensure reliable positioning.

\textit{2) Insider Threats:} UAVs frequently join, leave, and change roles based on task requirements, resulting in highly dynamic and self-organizing UAV networks \cite{Du2025Survey}. Traditional authentication mechanisms mainly rely on one-time initial identity verification, which are insufficient to defend against stealthy internal threats \cite{Ceviz2025Survey}. For instance, compromised UAVs may behave normally during initial authentication but later launch insider attacks such as task disruption and coordinated sabotage. As a result, dynamical trust management is essential to evaluate node trustworthiness based on behavioral and contextual information.

\textit{3) Multi-hop Penetrations:} Modern attacks against UAV swarms increasingly adopt multi-hop penetration strategies \cite{Ceviz2025Survey}. For instance, attackers may initially compromise a small number of UAVs and subsequently propagate malicious influence via inter-UAV links and cooperative interactions, forming stealthy attack paths across the swarm. The highly dynamic topology and complex interdependencies among UAVs make it challenging to timely trace attack origins and mitigate their propagation. Conventional security mechanisms lack the capability for intelligent attack reasoning and proactive system hardening.

To address these critical challenges, a new paradigm for securing UAV swarms is required. Such a paradigm should fully exploit the hierarchical and distributed characteristics of low-altitude wireless networks and enable coordinated security intelligence across cloud, network, and UAV layers. In this context, cloud-edge-end cooperative architectures \cite{Sun2026CFMMIMO,Zhao2026Joint} have emerged to provide complementary security capabilities, where the cloud performs global threat intelligence analysis, edge networks support real-time distributed protection, and UAV swarms execute autonomous sensing and defense, as illustrated in Fig.~\ref{fig:1}.
Building upon this architectural foundation, three key solutions are further developed to secure UAV swarms: (i) a cooperative perception mechanism to resist GPS spoofing via swarm-level information fusion and adaptive resource allocation; (ii) a continuous authentication method based on behavioral trust evolution to identify malicious insider UAVs; (iii) an intelligent attack forensics strategy to proactively trace multi-hop attack paths.

In this article, we present a comprehensive framework for secure collaborative UAV swarms in low-altitude wireless networks. We first introduce a cloud-edge-end cooperative architecture that enables scalable and resilient UAV swarm defense. We then analyze key security challenges and propose corresponding solutions, including cooperative spoofing defense, behaviora trust authentication, and intelligent attack forensics. These solutions collectively provide a holistic approach toward trustworthy and resilient UAV swarm operations in dynamic and adversarial environments.

The remainder of this paper is structured as follows. The cloud-edge-end cooperative architecture and related works are introduced in Section II. 
Section III discusses key security challenges in UAV swarms. Section IV introduces the proposed solutions. Section V provides performance evaluation results. Section VI outlines open research directions, and Section VII concludes the paper.

\section{Cloud-Edge-End Architecture for Secure Collaborative UAV Swarms}

\subsection{Architecture Design}
\begin{figure}[!t]
\setlength{\abovecaptionskip}{-0.03cm}\vspace{-3mm}
\centering
\includegraphics[width=1.02\linewidth]{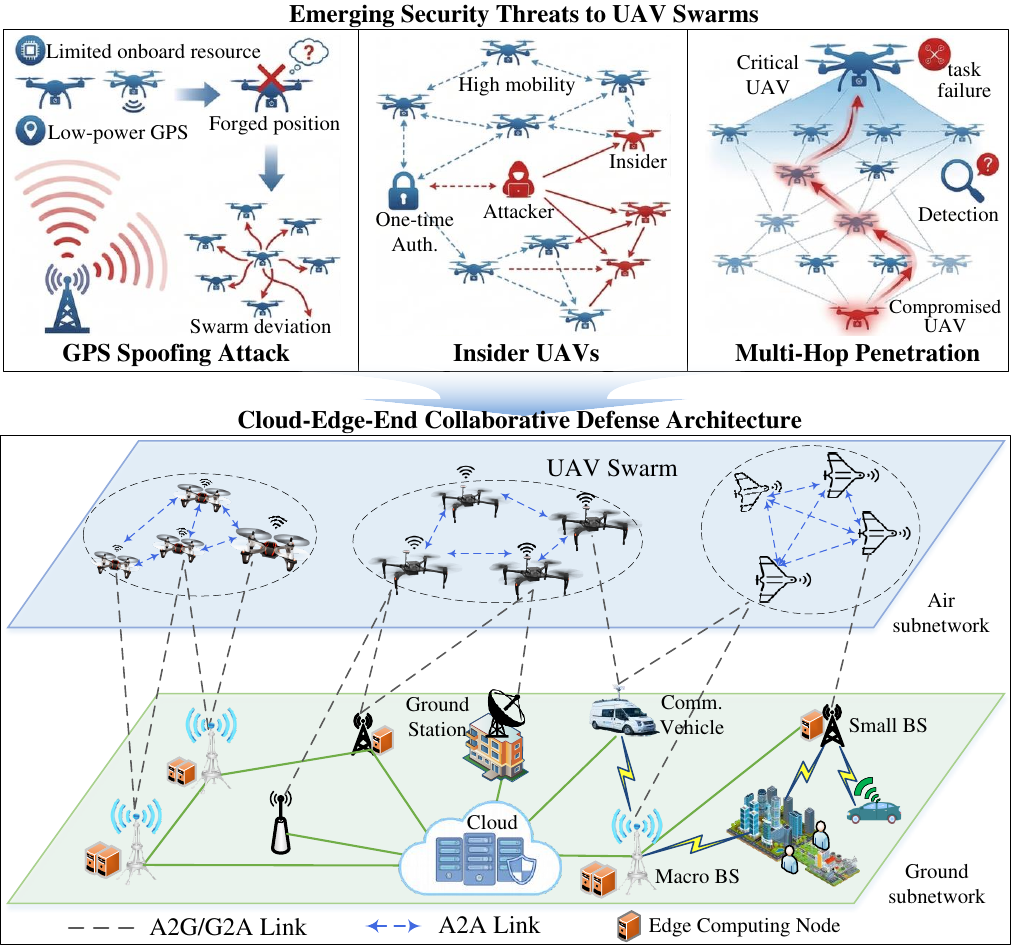}
  \caption{Overview of emerging security threats to UAV swarms and the cloud-edge-end collaborative defense architecture in low-altitude wireless
networks.}\label{fig:1}\vspace{-3mm}
\end{figure}

To address the fundamental challenges arising from highly dynamic network topologies and limited onboard security capabilities, we propose a cloud-edge-end collaborative architecture for UAV swarms in low-altitude wireless networks, as illustrated in Fig.~\ref{fig:1}. By leveraging the cloud's global intelligence, the edge's distributed protection capability, and the UAV swarm's autonomous decision-making and real-time sensing capability, the proposed architecture establishes a hierarchical, adaptive, and cooperative defense framework. 

\emph{1) Cloud} acts as the global intelligence and security orchestration center to secure UAV swarms. With abundant computing and storage resources, the cloud performs large-scale security analytics and global decision-making. \ding{172} By aggregating historical attack data, UAV operational logs, and multi-source monitoring information from the edge and UAV layers, the cloud conducts global threat modeling and cross-domain correlation analysis to establish comprehensive security situational awareness.
\ding{173} Based on global analysis, the cloud dynamically generates optimized defense policies and mitigation strategies, and disseminates security instructions to edge nodes and UAV swarms to ensure coordinated and consistent protection.

\emph{2) Edge networks} provide distributed computing and localized protection by deploying edge nodes across aerial and ground communication infrastructures, such as relay UAVs, ground stations, and cellular base stations (BSs). \ding{172} Edge nodes process UAV communication traffic locally, enabling low-latency anomaly detection, traffic inspection, and hierarchical data forwarding, thereby reducing the processing burden on the cloud.
\ding{173} Edge nodes can cooperatively detect UAVs' abnormal activities by continuously analyzing and monitoring UAVs' link conditions, control signaling, and behavioral patterns within their coverage areas.

\emph{3) UAV swarms} are equipped with onboard sensing, communication, and lightweight security modules for autonomous perception and collaborative tasks.
\ding{172} Each UAV performs real-time threat detection based on locally observed information for rapid anomaly detection.
\ding{173} UAVs can share threat intelligence, collaboratively detect suspicious nodes, and autonomously execute defense strategies via inter-UAV communications and distributed consensus, thereby facilitating self-organized swarm protection.


\subsection{State-of-The-Art Approaches for Securing UAV Swarms}
Recent studies have revealed significant security risks in UAV swarms from the perspectives of positioning, communications, and systems.
Regarding UAV positioning security, Sathaye \textit{et al.} \cite{Sathaye2022Experimental} developed a GPS signal generator to launch GPS spoofing attacks, which can takeover commercial UAVs (e.g., trajectory and speed) by carefully manipulating spoofing signals in real time. 
Chen \textit{et al.} \cite{Chen2022DPM} further proposed a UAV position manipulation attack, which can precisely manipulate flight trajectories to induce an invading UAV to a redirected destination.
To mitigate GPS spoofing, Sathaye \textit{et al.} \cite{Sathaye2022SemperFi} designed a single-antenna GPS receiver named SemperFi, which equips an extended Kalman filter module on UAVs to recover legitimate positioning signals under spoofing attacks. 
By comparing reported GPS positions and inter-UAV distance measurements in UAV swarms, Bi \textit{et al.} \cite{Bi2024Spoofing} modeled compromised UAV identification in collaborative GPS spoofing detection as a non-convex localization feasibility problem, which is solved efficiently using semidefinite relaxation. 

Regarding UAV communication security, Piana \textit{et al.} \cite{Piana2025Challenge} designed a physical-layer multi-round authentication protocol in UAV communications, where a verifier issues movement challenges and validates UAV authenticity by checking whether the measured channel gain matches expected statistics. A zero-sum game was also utilized to decide challenge positions.
Sikarwar \textit{et al.} \cite{Sikarwar2024SECURE} explored lightweight physically unclonable functions (PUFs) for efficient UAV-aided identity authentication with privacy preservation for ground vehicles. 
For secure UAV-assisted computation offloading, Chen \textit{et al.} \cite{Chen2025Blockchain} developed a hierarchical Stackelberg game framework to optimize resource scheduling in blockchain-enabled systems. 

Recent research has begun leveraging large language models (LLMs) to ensure UAV system security. Deng \textit{et al.} \cite{Deng2024PentestGPT} designed an LLM-empowered automated penetration testing framework named PentestGPT composed of multiple self-interaction modules to cooperatively support long reasoning processes in complex security analysis. 
Cao \textit{et al.} \cite{Cao2025LLM} studied a novel situational awareness framework based on the LLM agent in space-air-ground integrated networks, which is capable of multi-domain threat intelligence fusion and automatic defense strategy generation. 

Sect.~\ref{challenges} discusses key challenges in current UAV swarm security approaches, followed by potential solution in Sect.~\ref{solutions}.

\section{Challenges of UAV Swarms Security}\label{challenges}
This section highlights key challenges in securing UAV swarms.

\vspace{-2mm}
\subsection{From Single-UAV-Based Defense to Collaborative Defense Against GPS Spoofing with Free-Rider Resistance}\label{Challenge1}

UAVs heavily rely on GPS signals for navigation, positioning, and coordinated swarm operations. Due to their weak signal strength and open broadcast nature, GPS signals are highly vulnerable to spoofing attacks, which can be launched at low cost \cite{Sathaye2022Experimental}. Smart adversaries can inject forged GPS signals to gradually introduce location deviations (i.e., strategically dynamic GPS bias), misleading UAV navigation and potentially leading to mission failures. 

Traditional defense approaches mainly rely on multi-source sensor fusion on individual UAVs \cite{Sathaye2022SemperFi}, which are often ineffective under smart GPS spoofing attacks.
Although few existing recent works have explored collaborative GPS spoofing defense, they mainly operate under the optimistic assumption that all UAVs voluntarily participate. 
In practice, UAVs may behave as self-interested agents due to privacy concerns and limited onboard computation and energy resources. Consequently, some UAVs may act as \emph{free riders}, benefiting from the collective defense without contributing sensing or computation resources, thereby degrading the effectiveness of collaborative defense mechanisms.
Therefore, it is critical to develop incentive-compatible collaborative defense frameworks to encourage honest participation from UAVs while mitigating free-rider behaviors. 

\subsection{From One-Time Identity Authentication to Trust-Aware Behavioral Authentication Against Insider UAVs}\label{Challenge2}

UAV swarms typically operate in a task-driven and self-organizing manner, where UAV nodes dynamically join, leave, and change roles \cite{Du2025Survey}. Such dynamic and decentralized characteristics introduce security risks associated with insider malicious nodes. Existing authentication mechanisms mainly rely on cryptographic techniques to verify UAV identities during initial network access, but they are insufficient to detect insider threats.

In practice, UAVs may be captured, compromised, or infiltrated by adversaries after joining the swarm. Compromised UAVs can then launch insider attacks, such as falsifying sensing data, manipulating swarm coordination, or disrupting mission execution \cite{Ceviz2025Survey}. Moreover, insider UAVs may intentionally exhibit compliant behaviors in early stages to evade detection, and later initiate strategic attacks. 
Therefore, beyond one-time identity authentication, continuous trust evaluation is necessary to capture behavioral characteristics against insider UAVs. Furthermore, due to limited onboard resources, individual UAV observations may be insufficient for comprehensive trust assessment, necessitating the integration of multi-source swarm information. 



\begin{figure*}
\setlength{\abovecaptionskip}{-0.03cm}\vspace{-3mm}
\centering
\includegraphics[width=18cm]{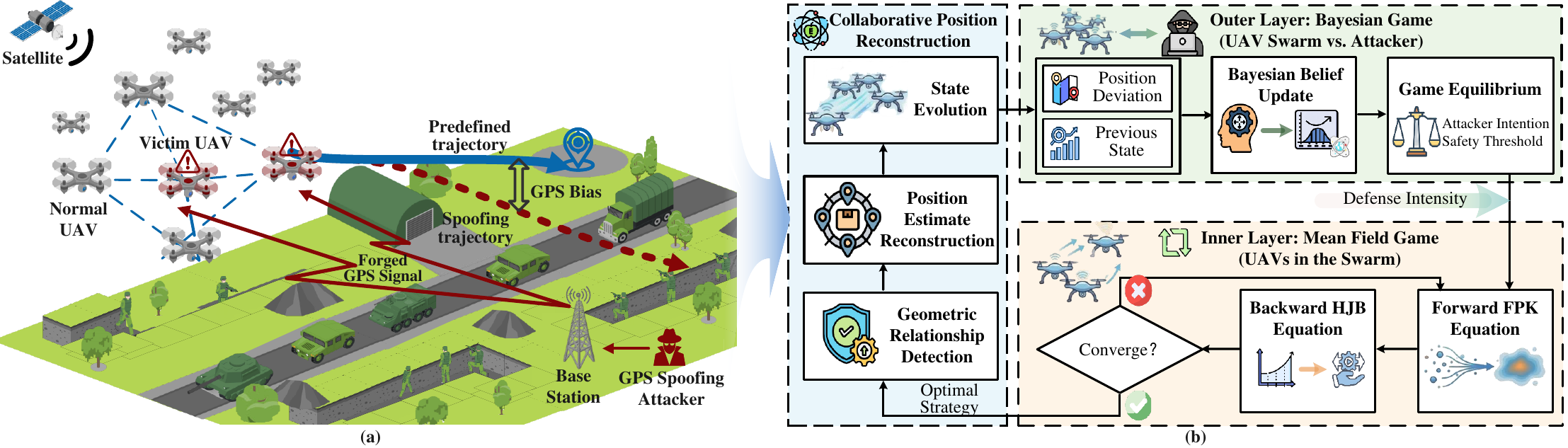}
\caption{(a) System model of cooperative GPS spoofing defense for UAV swarms, and (b) its workflow: 1) collaborative position reconstruction and 2) optimal attack-defense strategy based on the outer-layer Bayesian game (between the UAV swarm and the attacker) and inner-layer mean field game (among individual UAVs within the swarm).}
\label{fig:Fig2}\vspace{-1mm}
\end{figure*}

\subsection{From Passive Patching to Intelligent Attack Tracing Against Multi-Hop Penetrations}\label{Challenge3}

UAV swarms exhibit highly dynamic topologies due to UAV mobility and unstable communication links. Smart adversaries can exploit network vulnerabilities to launch multi-hop penetration attacks by progressively compromising UAV nodes and propagating malicious influence within the swarm. Such attacks typically involve multiple vulnerabilities and dynamically constructed attack paths, making them highly stealthy and difficult to detect.

Traditional vulnerability scanning and security patching methods mainly assume static network configurations. In practice, attackers can continuously adapt their strategies according to topology changes and system states in UAV swarms. Consequently, existing approaches have limited capability in reconstructing dynamically evolving attack paths in UAV networks in a timely manner.
Recent studies have explored LLMs for intelligent security analysis. However, applying LLM-based approaches to UAV swarms remains challenging. First, UAV swarm's configurations and vulnerability data often reach gigabyte scale, exceeding the context window capacity of current LLMs. Second, reasoning reliability remains a concern, as hallucinations and inference errors may lead to inaccurate attack analysis. 
Therefore, it is essential to develop intelligent multi-path attack reasoning mechanisms that can adapt to dynamic UAV network topologies and complex vulnerability correlations. 

\section{Solutions to Collaborative Defense of UAV Swarms}\label{solutions}
To address the above key security challenges in low-altitude UAV swarms, this section presents three complementary solutions: cooperative GPS spoofing defense, dynamic behavioral authentication, and proactive swarm protection. 

\subsection{Cooperative GPS Spoofing Defense for UAV Swarms}
As shown in Fig.~\ref{fig:Fig2}, this subsection develops a cooperative GPS spoofing defense framework consisting of (i) a collaborative position reconstruction method and (ii) a Bayesian-MFG double-layer decision mechanism.

\emph{1) Collaborative Position Reconstruction.} Since a single UAV often cannot reliably detect stealthy GPS spoofing, we develop a collaborative perception model based on multi-dimensional geometric residuals that exploits the spatial geometry of the UAV swarm for accurate position reconstruction. Each UAV first measures relative distances via line-of-sight (LoS) links to compute an edge residual metric \textit{k}, which quantifies the discrepancy between measured inter-UAV distances and GPS-reported positions. To filter low-confidence observations, a nonlinear mapping function converts these residuals into spoofing probabilities. When a UAV is suspected of being spoofed, a collaborative localization procedure is triggered. High-confidence neighboring UAVs are selected as anchor nodes, and semi-definite programming (SDP) \cite{Bi2024Spoofing} is used to construct linear constraints for estimating the victim UAV's true location. The reconstructed position is then fused with the original GPS estimate to generate a weighted corrected position. 

\emph{2) Optimal Attack-Defense Strategy Based on Bayesian-MFG Game.} In GPS spoofing attacks, the attacker can dynamically inject position biase \textit{a}
to maximize trajectory deviation while remaining undetected. Meanwhile, each resource-constrained UAV decides its collaborative defense intensity (CDI) \textit{p} (i.e., the collaborative communication range), which influences the number of neighboring cooperators, to minimize the expected cost \textit{C} consisting of communication latency, computational energy consumption, and position deviation risk. For efficient intra-swarm defense coordination, we design a double-layer decision mechanism. At the outer layer, a Bayesian game models the strategic interaction between the attacker and the UAV swarm. The attacker's utility \textit{U} is related to the GPS bias \textit{a} and the probability of being detected. Based on observed geometric residuals \textit{k}, the swarm updates its belief about the attacker’s behavior and dynamically adjusts the swarm-level defense intensity (i.e., the maximal CDI). 

Given this global defense strategy, an inner-layer mean field game (MFG) governs the collective defense behavior of individual UAVs. 
For each UAV, its state variable \textit{s(t)} is an estimate of its position deviation, and its goal is to find the optimal CDI strategy \textit{p*(t)} to minimize the expected cost \textit{C(t)}.
Specifically, the Hamilton-Jacobi-Bellman (HJB) equation determines the optimal CDI strategy for each UAV, while the Fokker-Planck-Kolmogorov (FPK) equation characterizes the evolution of the swarm state distribution under stochastic disturbances. 

To address the coupling between outer-layer strategic confrontation and inner-layer swarm dynamics, the outer-layer Bayesian equilibrium is approximated via a fictitious play (FP) method, while the inner-layer HJB-FPK equations are solved using an implicit finite-difference scheme with the Thomas algorithm. Such nested optimization is executed within a model predictive control (MPC) rolling-horizon framework, enabling the adjustment of swarm-level defense intensity at each time slot.



\begin{figure*}
\setlength{\abovecaptionskip}{-0.03cm}\vspace{-3mm}
\centering
\includegraphics[width=18cm]{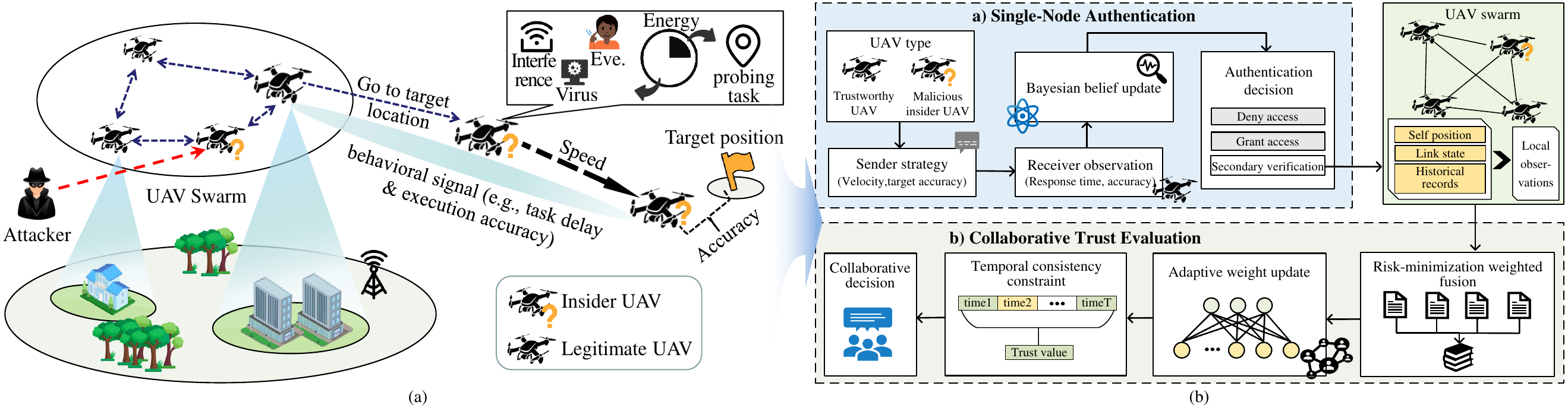}
\caption{(a) System model of dynamic behavioral authentication for UAV swarms, and (b) its workflow: 1) single-node behavioral authentication and 2) distributed collaborative trust evaluation.}
\label{fig:Fig3}\vspace{-1mm}
\end{figure*}

\subsection{Trust-Aware Behavioral Authentication for UAV Swarms}
This subsection develops a trust-aware behavioral authentication framework consisting of (i) a dynamic behavioral authentication model and (ii) a distributed collaborative trust mechanism to mitigate insider threats, as shown in Fig.~\ref{fig:Fig3}.

\emph{1) Dynamic Behavioral Authentication.}
In a UAV swarm, each node can be either a \textit{legitimate} UAV or an \textit{insider} adversary, while authentication decisions are made by the swarm manager based on UAVs' behavior signals. Particularly, the swarm manager mixes real tasks with lightweight probing tasks (e.g., rapid maneuvering to a designated location), and dynamically assigns them to the target UAV. Then, observable behavioral signals of the target UAV can be extracted from task delay and execution accuracy (i.e., the distance between its actual arrival location and mission target location). Note that UAV's energy consumption is commonly proportional to the square of its speed. 
Generally, legitimate UAVs tend to complete assigned tasks efficiently within normal energy budgets, whereas malicious insiders often conserve resources for later attacks, resulting in higher delays or reduced execution accuracy. 
Based on the behavioral signals in executing test tasks, the swarm manager periodically updates its posterior belief regarding the UAV type (i.e., legitimate or insider) through Bayes' rule. Then the swarm manager performs graded access control based on the predefined trust thresholds: highly credible nodes are admitted, low-credibility nodes are rejected, and nodes with uncertain credibility are forwarded to the following collaborative authentication stage. 

\emph{2) Distributed Collaborative Trust Evaluation.}
As insider UAVs may evade single-node authentication through temporary behavioral camouflage, a distributed swarm-level trust evaluation mechanism is further introduced. Specifically, multiple neighboring UAVs generate local trust sequences of the target node based on historical interactions. Here, continuous trustworthy interactions contribute to positively accumulated trust with temporal decay, while malicious behaviors trigger a betrayal penalty that amplifies distrust to discourage strategic camouflage. Finally, local trust assessments are fused through adaptive weighting and sliding-window aggregation to produce a joint trust score, which mitigates local observation noise and enhance swarm resilience against intermittent stealthy attacks. 

\begin{figure*}[t]\setlength{\abovecaptionskip}{-0.03cm}
  \centering
  \includegraphics[width=0.78\linewidth]{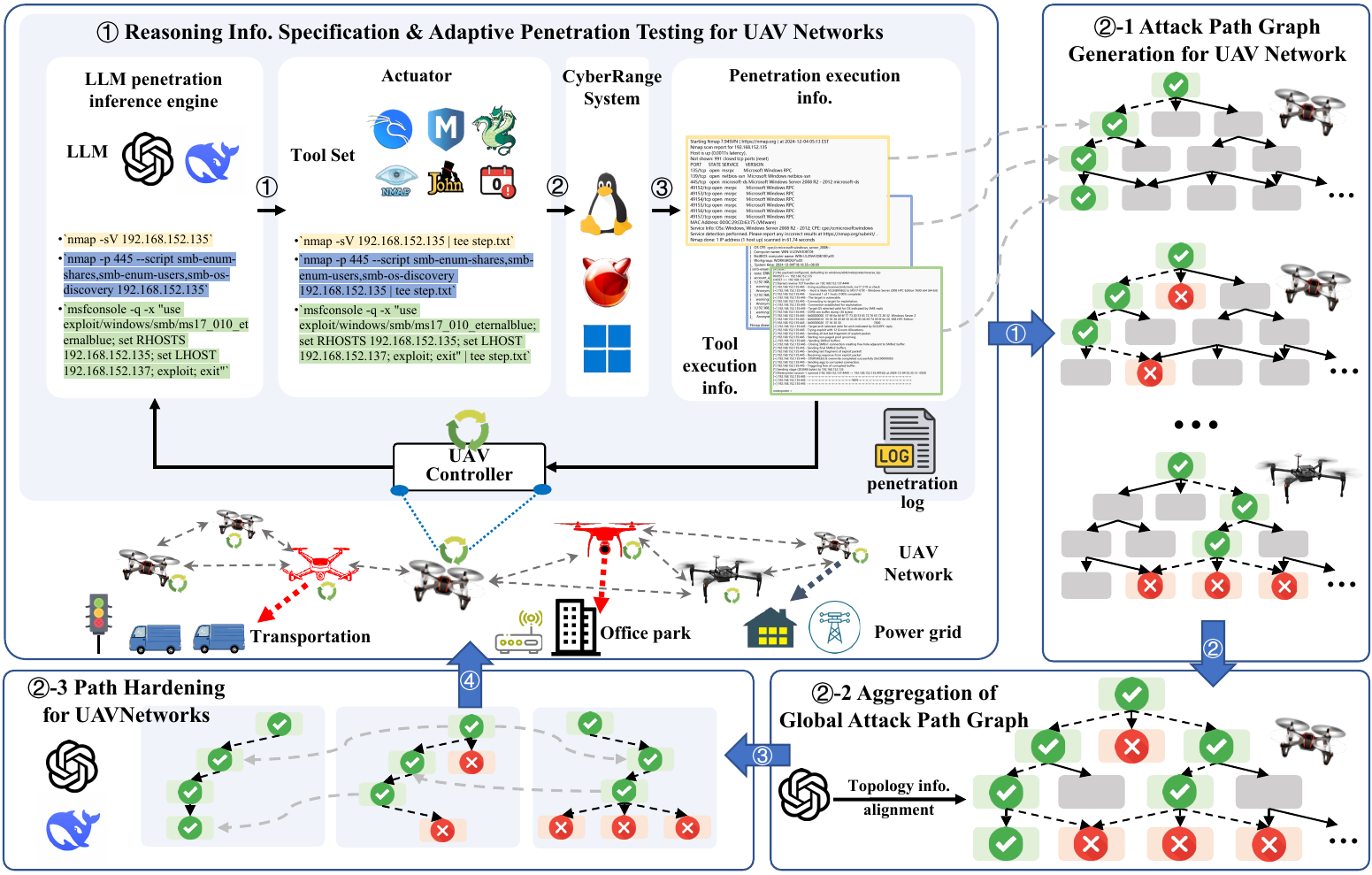}
    \caption{Illustration of adaptive proactive protection for UAV swarms based on multi-agents, including (i) reasoning information specification, and (ii) collaborative attack path tracing.}
  \label{Multi_agents}\vspace{-1mm}
\end{figure*}

\subsection{Multi-Agent Proactive Protection for UAV Swarms}
This subsection develops a proactive protection framework that includes (i) formal logic-based reasoning information specification and (ii) multi-agent cooperative attack path reasoning, as shown in Fig.~\ref{Multi_agents}.

\emph{1) Formal Logic-Based Reasoning Information Specification.}
To handle the massive and redundant volume of vulnerability scanning data within the UAV swarm, a formal logic-based specification method is designed. It systematically maps heterogeneous UAV network configurations and raw vulnerability reports into compact atomic facts. Using the Datalog logic framework, reasoning rules are defined based on Horn clause to describe the transitive relationships of potential attack behaviors across the UAV network. This specification process compresses large volumes of raw security data into structured reasoning information, significantly reducing computational complexity while preserving essential logical dependencies required for LLM-based penetration analysis.

\emph{2) Multi-Agent Collaborative Attack Path Tracing.}
As LLM-based single-agent systems often suffer from hallucinations and insufficient coverage in long-chain reasoning, a multi-agent collaborative architecture is developed. It dynamically schedules specialized agents with distinct roles, including path reasoning (PR), path deduplication  (PD), sub-path exploration  (SPE), and result verification (RV). 

The PR agent generates a set of potential attack paths based on the structured reasoning information obtained in phase 1. To ensure the self-consistency of any generated path, the RV agent initiates multiple verification instances in parallel and performs voting-based validation. To mitigate path redundancy, the PD agent identifies and merges logically equivalent attack steps via LLM, thereby refining the set of valid attack paths. The SPE agent runs a prefix decomposition-based exploration algorithm to improve attack graph coverage through recursive SPE.
Based on the traced attack graphs, the system determines the repair priority for each vulnerability according to its exploitation frequency across different attack paths and its topological proximity to critical assets, rather than relying on disruptive global patching. 

\section{Case Study}
We simulate a UAV swarm comprising 500 UAVs distributed following a 3D Poisson point process. Each UAV has a basic communication radius of 10 meters and a maximum sensing radius of 50 meters. To reflect UAV mobility, the routing links within the UAV swarm are dynamically updated.  
A forged ground station launches a GPS spoofing attack on an edge UAV within 2-5 seconds of the simulation window. By manipulating the time of arrival (ToA) and forging satellite signals, the attacker induces a continuous downward-diagonal trajectory drift of approximately 12 m/s.
In addition, a multi-hop penetration attack targeting core mission assets is injected between 10-30 seconds of the simulation window, during which the adversary performs lateral movement by exploiting distributed vulnerabilities across connected UAVs. Within the UAV swarm, 20\% of UAVs are assumed to be insider nodes that may misreport false positions during collaborative defense and leak private UAV configurations in multi-hop penetrations.

Six baseline strategies are adopted for comparison: (i) the Continuous Operation Strategy (COS), where each UAV maintains defense cooperation among all UAVs at all times; (ii) the Linear Feedback Strategy (LFS), where each UAV scales defense intensity proportionally to trajectory deviation; (iii) the Greedy Strategy (GS), where each UAV minimizes instantaneous overhead without considering long-term effects; (iv) a Formal Logic Strategy (FLS) that performs exhaustive state-space reasoning, (v) a Single-Agent Strategy (SAS) relying on sequential vulnerability reasoning, and (vi) a Greedy Patching (GP) strategy that immediately patches vulnerabilities upon detection.

\begin{figure}[!t]
    \centering
    \centering\setlength{\abovecaptionskip}{-0.1cm}\vspace{-3mm}
    \includegraphics[width=6.2cm,height=4.8cm]{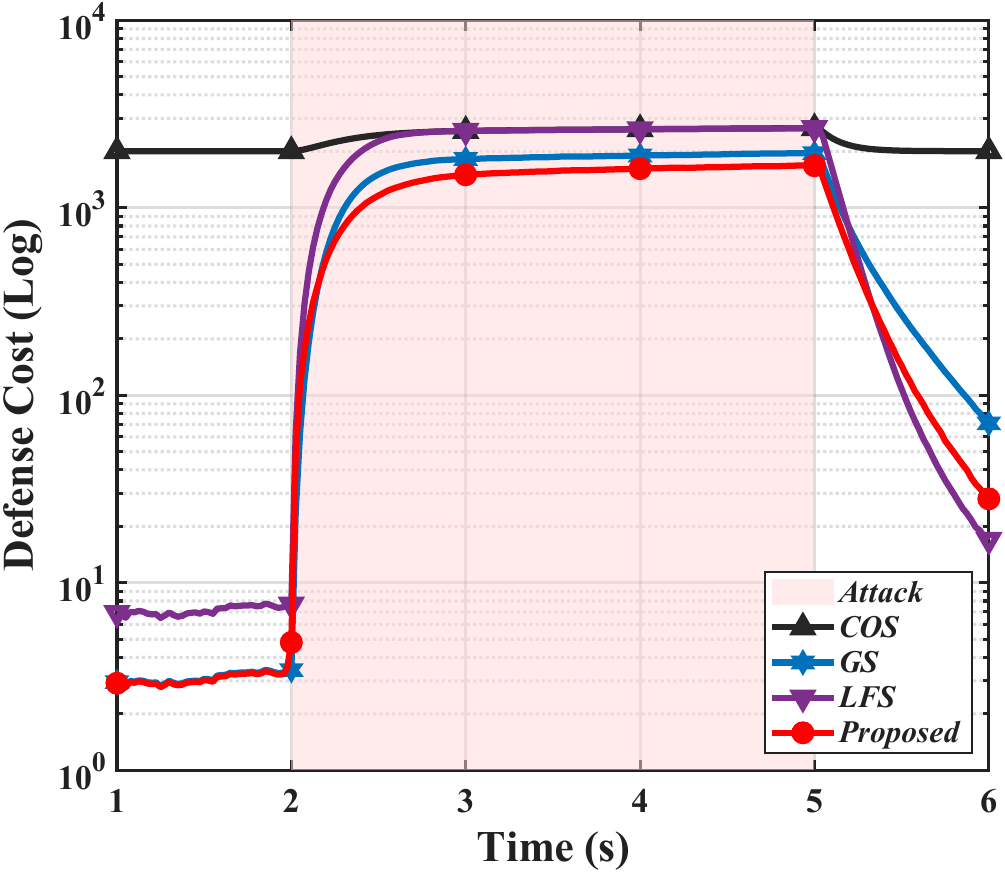}
    \caption{Evolution of instantaneous defense costs under GPS spoofing attacks in different strategies.}
    \label{evaluation:GPS}\vspace{-0mm}
\end{figure}

\begin{figure}[!t]
    \centering
    \centering\setlength{\abovecaptionskip}{-0.1cm}\vspace{-3mm}
    \includegraphics[width=9cm]{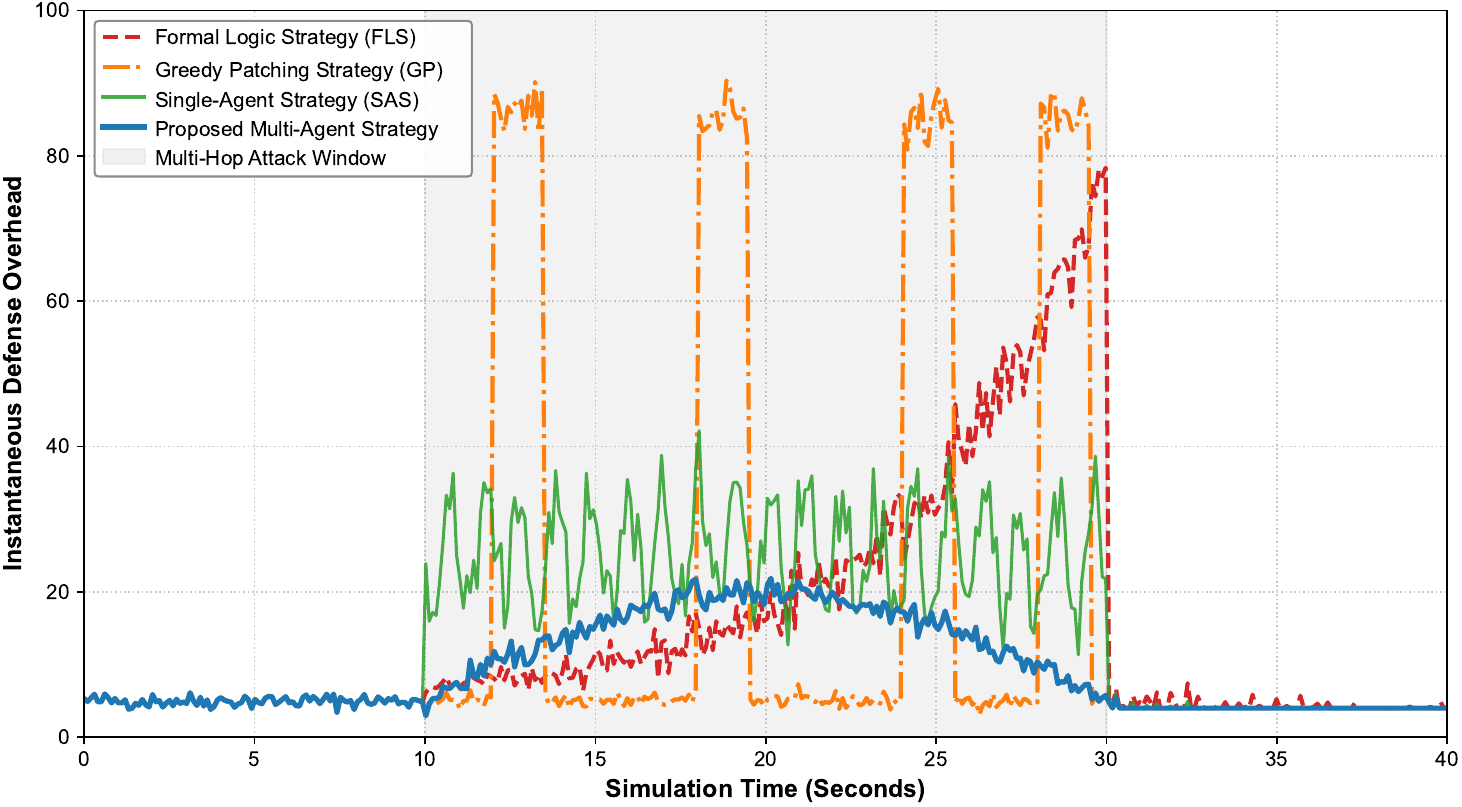}
    \caption{Evolution of instantaneous defense overhead under multi-hop penetration attacks in different strategies.}
    \label{evaluation:Hardening}\vspace{-2.5mm}
\end{figure}

Fig.~\ref{evaluation:GPS} shows the instantaneous defense cost against GPS spoofing attacks under four strategies. As seen in Fig.~\ref{evaluation:GPS}, the COS strategy always incurs the highest cost, since all UAVs maintain maximal-range cooperative defense with their neighbors. The LFS strategy tends to overreact to sudden trajectory deviation and cause higher defense cost due to its fixed feedback gain. The GS strategy suffers increased resource consumption once the attack occurs, as it lacks long-term coordination among UAVs. 
In contrast, the proposed framework attains the lowest defense cost against GPS spoofing (i.e., during 2-5 seconds in Fig.~\ref{evaluation:GPS}). The reasons are as below. First, swarm-level collaboration guided by the MFG effectively coordinates individual defensive efforts across UAVs, thereby preventing excessive resource consumption at individual UAVs. 
Second, the proposed trust-based behavioral authentication can effectively resist malicious insiders within the swarm, improving the overall honesty of participating UAVs. In addition, once the attack subsides, the system quickly returns to a low-power operating state, achieving an effective balance between security and efficiency.

Fig.~\ref{evaluation:Hardening} shows the instantaneous defense overhead during multi-hop penetrations under four strategies. As seen in Fig. \ref{evaluation:Hardening}, with the increase of attack path depth, the FLS strategy exhibits an exponential growth in defensive computation overhead. The SAS strategy produces highly oscillatory overheads, due to potential reasoning hallucinations of a single LLM agent and the subsequent generation of invalid verification scripts. The GP strategy results in severe overhead spikes, due to service interruptions caused by immediate patching of UAV nodes. Conversely, the proposed multi-agent collaboration strategy maintains the lowest average defense overhead. The reasons are as below. First, the formal specification mechanism effectively compresses the redundant reasoning space, while the trust model mitigates insiders within the swarm. Second, during the multi-hop penetration phase (i.e., 10-30 seconds in Fig.~\ref{evaluation:Hardening}), multiple LLM agents are dynamically scheduled to perform recursive sub-path exploration to identify critical vulnerabilities. Finally, via adaptive UAV network hardening, the system prioritizes the mitigation of high-risk attack paths, thereby repairing network-level vulnerabilities while maintaining overall swarm availability. 

\section{Future Directions}

\subsection{Distributed LLM Inference for UAV Swarms via Mixture-of-Experts}
LLMs empower low-altitude wireless networks with powerful reasoning and decision-making capabilities. However, due to high computation and memory demands, it remains prohibitive to directly deploy LLMs on resource-limited UAV platforms. A promising direction is to leverage mixture-of-experts (MoE) architectures that partition an LLM into multiple shards across UAVs and edge nodes, thereby facilitating distributed LLM inference. In this paradigm, UAVs can cache and update specialized expert models from nearby edge nodes, while each inference task is dynamically routed to the corresponding UAV that hosts the appropriate expert. 
However, key challenges remain in latency-aware expert selection and communication-efficient collaborative inference among UAVs.

\subsection{Cyber-Physical Agent Collaboration in Low-Altitude Intelligent Networks}
Next-generation low-altitude wireless networks will evolve into cyber-physical intelligent systems integrating embodied agents (e.g., physical UAVs) with virtual AI agents (e.g., digital twins and cloud-based service agents). 
To execute complex missions, both physical and virtual agents can dynamically publish and update their capabilities. Then suitable agent teams can be dynamically formed via capability-aware semantic matching and workflow orchestration.  
However, several challenges remain, including semantic alignment, real-time synchronization, and scalable agent
coordination among heterogeneous agents.


\subsection{Privacy-Preserving Collaborative Intelligence for UAV Swarms}
As UAVs frequently exchange sensitive sensing data, mission information, and operational states, privacy preservation becomes a critical issue in UAV swarms. For instance, shared observations and intermediate reasoning results during multi-agent collaboration may expose sensitive mission details or operational patterns. 
Existing studies on privacy preservation have explored distributed AI learning, LLM security guardrails, and differential privacy mechanisms. However, how to design lightweight privacy-preserving mechanisms for resource-constrained low-altitude agents, while achieving an effective balance between privacy and collaborative intelligence, remains a research challenge.

\section{Conclusion}
This paper has investigated the security challenges and potential solutions for enabling secure collaborative UAV swarms in low-altitude wireless networks. First, we have introduced a cloud-edge-end collaborative defense architecture to secure UAV swarms. Then, we have reviewed the state of the art, identified three critical security challenges across perception, communication, and network layers, and presented corresponding solutions, including cooperative GPS spoofing defense, dynamic behavioral trust authentication, and multi-agent proactive attack forensics. 
Finally, we have outlined several promising research directions, including distributed LLM intelligence, cyber-physical agent collaboration, and privacy-preserving swarm intelligence. By advancing secure collaborative mechanisms for UAV swarms, this work contributes to building trustworthy and resilient low-altitude wireless networks.





\end{document}